%% file: fair.tex
\documentclass[a4paper,11pt,titlepage,oneside,usenames,dvipsnames,svgnames,x11names,citecolor=NavyBlue,linkcolor=OliveGreen,urlcolor=blue]{article}
\pdfoutput=1 % if your are submitting a pdflatex (i.e. if you have
\usepackage{jheppub} % for details on the use of the package, please
\usepackage[T1]{fontenc} % if needed
\usepackage[usenames,dvipsnames,svgnames,x11names]{xcolor}

\usepackage{amsfonts}
\usepackage{amsmath}

\usepackage{amsthm}
\input{custom.tex}

\newcommand{\whZ}{\widehat Z}
\newcommand{\ket}[1]{\left\vert#1\right\rangle}
\newcommand{\bra}[1]{\left\langle#1\right\vert}
\newcommand{\braket}[1]{\langle #1\rangle}

\newcommand{\ketbra}[2]{\left\vert#1\right\rangle\!\!\left\langle#2\right\vert}
\DeclareMathOperator{\sech}{sech}

\usepackage[mathscr]{eucal}
\usepackage{comment}

\title{Proof of Renyi QNEC for free fermions}
\author{Pratik Roy}
\affiliation{Center for Quantum Information Theory of Matter and Spacetime, and\\ Center for Strings, Gravitation and Cosmology,\\ Department of Physics, Indian Institute of Technology Madras, Chennai 600036, India}
\emailAdd{roy.pratik92@gmail.com}

\abstract{Quantum null energy condition (QNEC) is usually stated as a bound on the expectation value of null components of the stress energy tensor at a point in terms of second null shape variations of the entanglement entropy at the same point. It can be recast as the statement that the sign of the second null shape variation of the relative entropy of any state with respect to the vacuum is positive. Using instead a Renyi generalization of relative entropy, called sandwiched Renyi divergence (SRD), leads to what is termed the Renyi QNEC: the second null shape variation of SRD of any state with respect to the vacuum is positive. In this work, we prove the Renyi QNEC for free and superrenormalizable fermionic quantum field theories in spacetime dimensions greater than 2 using null quantization, for the case where the Renyi parameter $n>1$. We end with comments on multiple possible generalizations.}

\begin{document} 
\maketitle
\flushbottom

%%%%%%%%%%%%%%%%%%%%%%%%%%%%%%%%%%%%%%%%%%%%%%%%%%%%%%%%%%%%%%%%%%%%%%%%%%%%%%%%%%%%%
%%%%%%%%%%%%%%%%%%%%%%%%%%%%%%%%%%%%%%%%%%%%%%%%%%%%%%%%%%%%%%%%%%%%%%%%%%%%%%%%%%%%%
%%%%%%%%%%%%%%%%%%%%%%%%%%%%%%%%%%%%%%%%%%%%%%%%%%%%%%%%%%%%%%%%%%%%%%%%%%%%%%%%%%%%%
\section{Introduction}

The holographic principle \cite{tHooft:1993dmi, Susskind:1994vu} (see also \cite{Bousso:2002ju}) and its explicit example, the AdS/CFT correspondence \cite{Maldacena:1997re, Witten:1998qj, Gubser:1998bc}, have revolutionized our understanding of (quantum) gravity over the past three decades. An entry of the AdS/CFT ``dictionary'', the HRRT formula \cite{Ryu:2006bv, Hubeny:2007xt}, provides a translation between the entanglement structure of quantum field theories and geometric properties of extremal surfaces in asymptotically AdS spacetimes. This has led to major advances in our understanding of the AdS/CFT correspondence, on both sides of the story.

An interesting development in recent years has been the quantum focussing conjecture \cite{Bousso:2015mna}, a generalization of the classical focussing theorem. The quantum focussing conjecture (QFC) reduces in a particular limit to what has been termed the quantum null energy condition (QNEC), which states that along null congruences with vanishing expansion and shear, one has
\begin{equation}			\label{eq:QNEC_0}
  \langle T_{kk}\rangle \geq \frac\hbar{2\pi}S'',
\end{equation}
where $T_{kk}=T_{ab}k^ak^b$, $T_{ab}k^ak^b$ is the stress-energy tensor of a quantum field theory (QFT) regarded as a quantum operator, $k^a$ is the tangent to the null congruence, and $S''$ denotes the second variation of the entanglement entropy (EE) as the entangling region is deformed along $k^a$ at one point. A more precise statement can be found in \cite{Bousso:2015wca}. We note that the QNEC does not involve the bulk gravitational constant, $G_N$, and as such is a statement purely about QFTs.

As discussed in \cite{Bousso:2015mna}, the variation $S''$ of the entanglement entropy appearing in QNEC is of the form
\begin{equation}
  S'' = \lim_{y'\to y}\frac{\delta^2S}{\delta\l(y)\delta\l(y')},
\end{equation}
with $\l$ the affine parameter along the geodesics of the null plane. Entanglement entropy of subregions is known to be universally UV divergent in QFTs (see e.g. \cite{Witten:2018zxz}), and one would like to have a better formulation of the QNEC. This is in fact easily achieved by rewriting (\ref{eq:QNEC_0}) in terms of the relative entropy \cite{Leichenauer:2018obf}. For density matrices associated with a region $R$, we have
\begin{equation}			\label{eq:rel_ent_defn}
  S_{\rm rel}(\r_R|\s_R) = \tr\r_R\log\r_R - \tr\r_R\log\s_R = \D\langle K_R^\s\rangle - (\D S)_R,
\end{equation}
where $K^\s=-\log\s$ is the modular Hamiltonian associated to a density matrix $\s$.\footnote{Note that here and in (\ref{eq:deltaK}) below, $\D x$ denotes a change in the quantity $x$, as opposed to Section \ref{sec:SRD}, where $\D$ is used to denote the modular Hamiltonian.} For arbitrary cuts of a null plane, one has the result that \cite{Casini:2017roe}
\begin{equation}
  K_R^\s = 2\pi\int_{V(y)}^\I (v-V(y))T_{vv}(y)dvd^{d-2}y,
\end{equation}
where $v$ is the null direction along the null surface, $y$ are the transverse directions, and $V(y)$ is an arbitrary curve on the null plane describing the choice of the cut. Choosing $\s_R$ to be the vacuum state, one can define a vacuum subtracted modular Hamiltonian,
\begin{equation}            \label{eq:deltaK}
	\D\braket{K} = 2\pi\int_{V(y)}^\I (v-V(y))\braket{T_{vv}(y)}dvd^{d-2}y.
\end{equation}
Taking two variations along $v$, and using the fact that the entanglement entropy for the vacuum state is stationary for null cuts, it is straightforward to see that (\ref{eq:QNEC_0}) is equivalent to
\begin{equation}			\label{eq:QNEC_rel_ent}
	\lim_{y'\to y} \frac{\d^2S_{\rm rel}(\r_R|\s_R)}{\d\l(y)\d\l(y')} \geq 0.
\end{equation}
Positivity of the off-diagonal components ($\l\neq\l'$) of the second variation of $S_{\rm rel}$ follows from strong subadditivity \cite{Bousso:2015wca}. However, we only focus on the ``diagonal'' terms of the variation, i.e., the limit $y'\to y$, throughout the work.

Evidence for validity of QNEC has been accumulating over the years. \cite{Bousso:2015wca} gave the first proof of QNEC for free and superrenormalizable bosons based on replica trick calculations. \cite{Malik:2019dpg} later generalized this proof to the case of free fermions. \cite{Koeller:2015qmn} proved QNEC for CFTs with holographic duals and their relevant deformations. \cite{Balakrishnan:2017bjg} proved the QNEC for general states of a CFT on Minkowski space using properties of modular Hamiltonians under shape deformations and causality. Finally, \cite{Ceyhan:2018zfg} provided a rigorous proof of QNEC for cases where the entangling region is a null cut in a general Poincare invariant QFT, using Tomita-Takesaki theory and the theory of half-sided modular inclusions.

Relative entropy is a measure of the distinguishability of a state $\r$ given a state $\s$. It is defined to be always positive, $S_{\rm rel}(\r|\s)\geq0$, and is monotonic under CPTP maps, $S_{\rm rel}(\F\r|\F\s)\leq S_{\rm rel}(\r|\s)$ \cite{Wehrl:1978zz, Witten:2018zxz}.\footnote{It is in fact monotonic under positive maps \cite{Muller:2015yot}.} This monotonicity property is often referred to in the literature as the data processing inequality (DPI). For states that are close to each other, DPI can be thought of as a constraint on the sign of the first derivative of $S_{\rm rel}$. QNEC is a constraint on the sign of the second derivative of $S_{\rm rel}$ in local quantum physics. Given the purely information-theoretic nature of QNEC (\ref{eq:QNEC_rel_ent}) and the amount of supporting evidence, it is natural to ask if generalizations of the relative entropy that are positive and monotonic also satisfy such constraints on their second derivatives.

One such measure is the sandwiched Renyi divergence \cite{Muller:2013Onqr, Wilde:2013bdg}, a Renyi generalization of the relative entropy. \cite{Lashkari:2014yva} showed that some sandwiched Renyi divergences (SRDs) can be written as correlation functions in quantum field theory. Based on the path integral expression for SRD in \cite{Lashkari:2014yva}, \cite{Lashkari:2018nsl} put constraints on correlation functions in QFT based on some known properties of the divergence. It was further conjectured there that the second null variation of the SRD should also be positive, thus providing a Renyi generalization of QNEC, which we will refer to as Renyi QNEC. A few examples where the conjecture holds were also demonstrated. Very recently, \cite{Moosa:2020jwt} gave a proof of the Renyi QNEC for free and superrenormalizable bosons in spacetime dimensions $D>2.$

In this work, we provide a proof of Renyi QNEC for free and superrenormalizable fermions in $D>2$ spacetime dimensions. Our proof closely follows the one presented in \cite{Moosa:2020jwt}. We use the formalism of null quantization to write an arbitrary state as an expansion around the vacuum, and reduce the Renyi QNEC for arbitrary states to a statement regarding a state perturbatively close to the vacuum. One can then evaluate the relevant SRD variations using expansions known in the literature, and show that the conjecture indeed is true in some cases. We note that the conjecture is not true for some values of the Renyi parameter, as also found in \cite{Moosa:2020jwt}. We do not investigate these.

We now present an outline of the rest of the paper. Section \ref{sec:SRD} begins with a review of the definition and properties of sandwiched Renyi divergence followed by a discussion of the Renyi QNEC conjecture. Section \ref{sec:setting} provides details of the set-up where we perform our computations. We briefly review the formalism of null quantization. Renyi QNEC is then reformulated as a perturbative statement, simplifying the calculations significantly. In Section \ref{sec:proof}, we proceed to prove Renyi QNEC, doing it in two ways, once for integer values of the Renyi parameter, and then for general values. The two methods provide complementary insight. Finally, we conclude with a discussion of possible generalizations and applications of Renyi QNEC. Two appendices contain details of the fermionic theory that we consider, and the calculation of correlation functions needed for completing the proof.

%%%%%%%%%%%%%%%%%%%%%%%%%%%%%%%%%%%%%%%%%%%%%%%%%%%%%%%%%%%%%%%%%%%%%%%%%%%%%%%%%%%%%
%%%%%%%%%%%%%%%%%%%%%%%%%%%%%%%%%%%%%%%%%%%%%%%%%%%%%%%%%%%%%%%%%%%%%%%%%%%%%%%%%%%%%
%%%%%%%%%%%%%%%%%%%%%%%%%%%%%%%%%%%%%%%%%%%%%%%%%%%%%%%%%%%%%%%%%%%%%%%%%%%%%%%%%%%%%
\section{Sandwiched Renyi divergence and Renyi QNEC}		\label{sec:SRD}

We begin this section by defining sandwiched Renyi divergence for finite dimensional quantum systems, in terms of density matrices. We then provide the basic definitions of modular theory, to be able to define SRD in QFTs in general. As discussed in \cite{Lashkari:2014yva, Lashkari:2018nsl}, sandwiched Renyi divergence for integer $n>1$ can be expressed in terms of one-sheeted Euclidean $2n$-point correlation functions. One can then evaluate the SRD explicitly and check the statement of (\ref{eq_renyi_qnec}), at least in free field theories. We provide a brief introduction to these ideas to finish this section.

%%%%%%%%%%%%%%%%%%%%%%%%%%%%%%%%%%%%%%%%%%%%%%%%%%%%%%%%%%%%%%%%%%%%%%%%%%%%%%%%%%%%%
%%%%%%%%%%%%%%%%%%%%%%%%%%%%%%%%%%%%%%%%%%%%%%%%%%%%%%%%%%%%%%%%%%%%%%%%%%%%%%%%%%%%%
\subsection{Preliminaries}

A quantum version of Renyi relative entropy, termed sandwiched Renyi divergence (SRD), was proposed for Type I von Neumann algebras in \cite{Muller:2013Onqr} and in \cite{Wilde:2013bdg}, see also Section 3.3 of \cite{Jaksic:2012}. States in these systems can be described in terms of density matrices. For two states described by density matrices $\r,\s,$ sandwiched Renyi relative $\a$-entropy\footnote{Alternatively called sandwiched Renyi divergence and Renyi relative entropy.} of $\rho$ with respect to $\s$ is defined to be
\begin{equation}
	S_\a(\r|\s) = \frac1{\a-1}\log\tr\lB\lb\s^{\frac{1-\a}{2\a}}\r\s^{\frac{1-\a}{2\a}} \rb^\a \rB, \qquad \a>0,\ \a\neq1,
\end{equation}
if the support of $\r$ is contained in the support of $\s$, otherwise, $S_\a(\r|\s)=\I.$ For $\a\in[1/2,\I)$, $S_\a(\r|\s)$ was shown to satisfy the DPI \cite{Frank:2013yoa, Beigi:2013drd}. The limit $\a\to1$ corresponds to Umegaki's relative entropy defined in (\ref{eq:rel_ent_defn}).

Our interest is in the Renyi relative entropy for QFTs, which correspond to Type III von Neumann algebras \cite{Araki:1964,Hislop:1981uh,Longo:1982zz,Fredenhagen:1984dc}. There is no notion of a density matrix for general von Neumann algebras, and the above definitions of relative entropy and Renyi relative entropy are not useful. It becomes essential to use the language of algebraic QFT, in particular, the theory of Tomita and Takesaki \cite{Tomita:1967, Takesaki:1970}. See \cite{Witten:2018zxz} for an accessible introduction to these ideas. We briefly review the basics needed to define SRD in QFT. The reader is also referred to \cite{Beny:2015} for a quick introduction to, and to \cite{Jaksic:2012} for a detailed treatment of, the algebraic approach in the finite dimensional context.

For an open set $R$ in $D$-dimensional Minkowski space, $M_D$, we consider the local algebra $\cA_R$ of operators supported in $R.$ Denote the vacuum state by $\W.$ States formed by acting with a finite number of operators on the vacuum comprise the vacuum sector, $\cH_0$, of the Hilbert space. A state  $\Psi\in\cH_0$ is called {cyclic} for $\cA_R$ if the states $A\ket\Psi,A\in\cA_R,$ are dense in $\cH_0$. The state $\Psi$ is called {separating} for $\cA_R$ if for $A\in\cA_R$, $A\ket\Psi=0\Rightarrow A=0.$ The Reeh-Schlieder theorem implies that the vacuum is a cyclic separating vector for the algebra associated to any subregion. Including what are known as weak limits of sequences of operators gives a closed algebra of operators, and we denote the corresponding Hilbert space by $\cH.$

Let $\Psi\in\cH$ be a cyclic separating state for $\cA_R$. The {Tomita operator} (for the state $\Psi$) is the antilinear\footnote{For an antilinear operator $W$, its adjoint for states $\l,\chi$ is defined by 
\[ \braket{\l\vert W\chi} = \overline{\braket{W^\dagger\l\vert\chi}}=\braket{\chi\vert W^\dagger\l}. \]
If $W$ is antiunitary, then $\braket{W\l\vert W\chi}=\braket{\chi\vert\l}.$} 
operator defined by $S_\Psi A\ket\Psi=A^\dagger\ket\Psi\A A\in\cA_R.$\footnote{One has to take a closure for the Tomita operator to be completely well-defined. We consider only the closed operator throughout.} $S_\Psi$ is invertible and has a unique polar decomposition as
\begin{equation}
	S_\Psi = J_\Psi\D_\Psi^{1/2},
\end{equation}
where $J_\Psi$ is an antiunitary called the {modular conjugation operator}, and $\D_\Psi^{1/2}$ is Hermitian and positive-definite; $\D_\Psi$ is called the {modular operator}.

Now, let $\Phi\in\cH'$ be another state, with $\cH'$ not necessarily the same as $\cH\ni\Psi.$ Let $\cA_R$ be an algebra that acts on both $\cH,\cH'$. The {relative Tomita operator} is defined by
\begin{equation}
	S_{\Psi\vert\Phi}:\cH\to\cH', \qquad S_{\Psi\vert\Phi}A\ket\Psi = A^\dagger\ket\Phi,
\end{equation}
where $\braket{\Psi|\Psi}=\braket{\Phi\vert\Phi}=1$, and one again needs to take a closure. The relative Tomita operator satisfies $S_{\W\vert\F}S_{\F\vert\W}=1$ and $S_{\W\vert\F}^\dagger S_{\F\vert\W}^\dagger=1$. The {relative modular operator} is defined to be
\begin{equation}
	\D_{\Psi\vert\Phi} = S_{\Psi\vert\Phi}^\dagger S_{\Psi\vert\Phi},
\end{equation}
which is unbounded and positive semi-definite. It is positive definite iff $S_{\Psi\vert\Phi}$ is invertible. The polar decomposition of the relative Tomita operator takes the form $S_{\Psi\vert\Phi} = J_{\Psi\vert\Phi}\D_{\Psi\vert\Phi}^{1/2}$. We note that the modular operator acts as, e.g., 
\begin{equation}
	\braket{B\Psi|\D_{\Psi\vert\Phi}|A\Psi} = \braket{A^\dagger\Phi|B^\dagger\Phi},
\end{equation}
which can be alternatively written as $B\D_{\Psi\vert\Phi}A=AB.$ The inverse of the relative modular operator satisfies $\D_{\Psi\vert\Phi}^{-1}=S_{\F\vert\Psi}S_{\F\vert\Psi}^\dagger.$ The relative modular operator comprises the core component of Tomita-Takesaki theory.

The relative modular operator can be used to define relative entropy in QFT  \cite{Araki:1976zv} as
\begin{equation}
	S_{\rm rel}(\Psi\vert\Phi) = -\langle\Psi\vert\log\Delta_{\Psi\vert\Phi}\vert\Psi\rangle.
\end{equation}
This reduces to (\ref{eq:rel_ent_defn}) for von Neumann algebras of Type I. For regions $\tilde R\subset R$, so that $\cA_{\tilde R}\subset \cA_R,$ the relative modular operators associated with given regions satisfy the monotonicity property: $\Delta_{\Psi\vert\Phi}^{R}\leq \Delta_{\Psi\vert\Phi}^{\tilde R}$\footnote{This means that the operator $\Delta_{\Psi\vert\Phi}^{\tilde R}-\Delta_{\Psi\vert\Phi}^{R}$ is a positive operator on $\tilde R$.}. This can be used to prove the monotonicity of $S_{\rm rel}(\Psi\vert\Phi)$ under restrictions of subsets, $S_{\rm rel}(\Psi\vert\Phi)(R)\geq S_{\rm rel}(\Psi\vert\Phi)(\tilde R)$.

%%%%%%%%%%%%%%%%%%%%%%%%%%%%%%%%%%%%%%%%%%%%%%%%%%%%%%%%%%%%%%%%%%%%%%%%%%%%%%%%%%%%%
%%%%%%%%%%%%%%%%%%%%%%%%%%%%%%%%%%%%%%%%%%%%%%%%%%%%%%%%%%%%%%%%%%%%%%%%%%%%%%%%%%%%%
\subsection{SRD for QFT and Renyi QNEC}

Sandwiched Renyi divergence between states $\F$ and $\Psi$ is defined by the relations \cite{Araki:1981mas,Berta:2016vnw,Lashkari:2018nsl}
\begin{equation}			\label{eq_srd_defn}
	\begin{split}
		S_p^R(\F|\Psi) =&\ \frac p{p-1}\sup_{\ket\chi\in\cH}\log\braket{\F|(\D_{\chi\vert\Psi})^{\frac1p-1}|\F}, \qquad p\in(1,\I)	\\
		S_p^R(\F|\Psi) =&\ \frac p{p-1}\inf_{\ket\chi\in\cH}\log\braket{\F|(\D_{\chi\vert\Psi})^{\frac1p-1}|\F}, \qquad p\in\big[{{\frac{_1}{^2}}},1).
	\end{split}
\end{equation}
This definition is a generalization of the $p$-norm of a matrix to unbounded operators.\footnote{For $p\geq1,$ the {(Schatten) $p$-norm} of a matrix $A$ is defined to be $|A|_p=\Big[\Tr\big(\sqrt{A^\dagger A}^{\,p}\big) \Big]^{1/p}$.} For $p>1,$ SRD is defined to be infinite if the vector $\Phi$ is not in the intersection of the domains of $\Delta_{\chi\vert\Psi}^{-1}$ for all $\ket\chi\in\cH.$ For $p<1,$ SRD is finite if $\Psi$ is cyclic separating \cite{Moosa:2020jwt}. 

SRD is non-negative and satisfies the data processing inequality, i.e., it monotonically decreases for algebras of smaller subregions \cite{Berta:2016vnw,Jencova:2016ire,Jencova:2017yie,Lashkari:2018nsl}. This makes it an interesting quantity to study from an information theoretic perspective. SRD is also monotonic in $p$, $S_p^R(\F|\Psi)>S_q^R(\F|\Psi)$ for $p>q\in[\frac12,1)\cup(1,\I)$. In the limit $p\to1$, $S_p^R(\F|\Psi)\to S_{\rm rel}(\Phi\vert\Psi)$.

Based on the similarities between the relative entropy and the sandwiched Renyi divergence, \cite{Lashkari:2018nsl} conjectured that its second variation should also be positive as it is for relative entropy,
\begin{equation}			\label{eq_renyi_qnec}
	\lim_{y'\to y}\frac{\delta^2 S_p^R(\F|\Psi)}{\d\l(y)\d\l(y')}\geq0,
\end{equation}
where the variation being considered is the same as in (\ref{eq:QNEC_rel_ent}). We note that we are still only concerned with the ``diagonal'' component of the variation. This was termed the Renyi QNEC conjecture in \cite{Moosa:2020jwt}, where a proof for the case of free (and super-renormalizable) bosons was provided. The goal of this paper is to extend their proof to the case of free (and super-renormalizable) fermions.

%%%%%%%%%%%%%%%%%%%%%%%%%%%%%%%%%%%%%%%%%%%%%%%%%%%%%%%%%%%%%%%%%%%%%%%%%%%%%%%%%%%%%
%%%%%%%%%%%%%%%%%%%%%%%%%%%%%%%%%%%%%%%%%%%%%%%%%%%%%%%%%%%%%%%%%%%%%%%%%%%%%%%%%%%%%
\subsection{Correlation functions from path integrals}		\label{subsec:corrfns}

Let us write $D$-dimensional Minkowski space as $M_D$, with metric $ds^2=-dt^2+dx^2+d\vec y\cdot d\vec y$. Let $\S$ be the initial value surface $t=0$, which we divide into two half-spaces, $x>0$ and $x<0$, denoted $V_{R,L}$. Their respective domains of dependence will be denoted $U_{R,L}$ with the corresponding local algebras being $\cA_{R,L}$. Let $\W$ be the vacuum of a QFT on $M_D$. States in this QFT can be prepared by inserting appropriately smeared operators in the path integral over the lower half of Euclidean time, $\t\leq0.$ We will only be interested in the set of states that have finite SRD with respect to the vacuum.

The modular operator on the vacuum state, $\D_\W$, leaves the vacuum invariant. However, positive powers, $\D_\W^\a,\ \a\in(0,1/2]$ take operators in $\cA_{R}$ and rotate them to the location $\q=-2\pi\a$ in the path integral over $\t\leq0$ \cite{Witten:2018zxz}. Here, $(r,\q)$ are polar coordinates on the $(\t,x)$-plane, with $z=x+i\t=re^{i\q}$.  Consider then an operator $\cO_{R}\in\cA_{R}$, and construct the excited states
\begin{equation}				\label{eq_Phi_state}
	\ket\F=\D_\W^{\q/2\pi}\cO_{R}(r,0)\ket\W = \cO(r,\q)\ket\W,
\end{equation}
with $\q\in[0,\pi]$, where we have used $\D_\W^{\q/2\pi}\cO_{R}(r,0)\D_\W^{-\q/2\pi}=\cO(r,\q)$. One can bound the SRD, $S_p^R(\F\vert\W)$, for $1\leq p\in\R,$ by choosing $\chi=\W$ in the supremum in (\ref{eq_srd_defn}),
\begin{equation}
	S_p^R(\F\vert\W) \geq  \frac p{p-1}\log\braket{\F|(\D_{\W\vert\W})^{\frac1p-1}|\F} = \frac p{p-1}\log||(\D_\W)^{\frac\q{2\pi}+\frac1{2p}-\frac12}\cO_R(r,0)\ket\W||^2.
\end{equation}
States of the form $\D_\W^\a\cO_R\ket\W$ generically have infinite norm for $\a>1/2$, which implies that $S_p^R(\F\vert\W)$ diverges for $\q<\pi-\pi/p$ and for $\q>2\pi-\pi/p.$ On the other hand, as discussed in \cite{Lashkari:2014yva, Lashkari:2018nsl}, sandwiched Renyi divergence for integer $p=n>1$ can be expressed in terms of one-sheeted Euclidean $2n$-point correlation functions, which are known to be finite. An overview of their construction follows.

The vacuum density matrix of the right half-space, $\w$, is given by a path-integral over the whole $\t$-plane, with cuts at $\t=0^\pm,x>0.$ Let $\f$ be the density matrix with operator insertions of $\F^\dagger,\F$ at $\pm(\pi-\q)$. For $\q\leq\pi/n$, with integer $n>1$, the operator $\w^{\frac1{2n}-\frac12}\f\,\w^{\frac1{2n}-\frac12}$ has a path integral representation as a wedge of opening angle $2\pi/n$ with two operator insertions. Sewing together $n$ such wedges evaluates $\tr\lB \lb \w^{\frac1{2n}-\frac12}\f\,\w^{\frac1{2n}-\frac12} \rb^n \rB$ as a $2n$-point correlation function of $\F,\F^\dagger$, with operators inserted at $z_k^\pm=re^{i\lb\frac{2\pi k}n\pm\q \rb}$ for $k=0,\ldots,n-1,$
\begin{equation}
	\tr\lB \lb \w^{\frac1{2n}-\frac12}\f\,\w^{\frac1{2n}-\frac12} \rb^n \rB = \left\langle\prod_{k=0}^{n-1}\F^\dagger(z_k^+)\F(z_k^-)\right\rangle,
\end{equation}
where the other directions $\vec y$ are suppressed. Then, for integer $n>1$, we can evaluate SRD for this configuration of fields with respect to the vacuum as
\begin{equation}
	S_n^R(\F|\W) = \frac1{n-1}\log\lb \frac{\langle\prod_{k=0}^{n-1}\F^\dagger(z_k^+)\F(z_k^-)\rangle}{\langle\F^\dagger(z_0^+)\F(z_0^-)\rangle^n} \rb.
\end{equation}

\begin{comment}
\begin{figure}
	\includegraphics[scale=0.4]{path_integral}
	\caption{abc}
\end{figure}
\end{comment}

This relation between SRD and correlation functions was used in \cite{Lashkari:2018nsl} to put constraints on correlation functions in QFT from known properties of SRD. It will be very useful for us too.

As a final point, note that although we defined SRD for QFTs using the abstract modular theory of Tomita and Takesaki, we have here resorted to using density matrices, and will continue to do so in what follows. This implicitly assumes that one is working with regularized entropy and energy-momentum tensor using an appropriate renormalization scheme. See \cite{Wall:2011hj} for a detailed discussion. It has recently been shown that this assumption holds in various situations \cite{Witten:2021unn, Chandrasekaran:2022cip, Chandrasekaran:2022eqq, Jensen:2023yxy, AliAhmad:2023etg, Klinger:2023tgi}\footnote{I thank Ro Jefferson for bringing \cite{AliAhmad:2023etg, Klinger:2023tgi} to my attention.}.

%%%%%%%%%%%%%%%%%%%%%%%%%%%%%%%%%%%%%%%%%%%%%%%%%%%%%%%%%%%%%%%%%%%%%%%%%%%%%%%%%%%%%
%%%%%%%%%%%%%%%%%%%%%%%%%%%%%%%%%%%%%%%%%%%%%%%%%%%%%%%%%%%%%%%%%%%%%%%%%%%%%%%%%%%%%
%%%%%%%%%%%%%%%%%%%%%%%%%%%%%%%%%%%%%%%%%%%%%%%%%%%%%%%%%%%%%%%%%%%%%%%%%%%%%%%%%%%%%
\section{Setting up}				\label{sec:setting}

%%%%%%%%%%%%%%%%%%%%%%%%%%%%%%%%%%%%%%%%%%%%%%%%%%%%%%%%%%%%%%%%%%%%%%%%%%%%%%%%%%%%%
%%%%%%%%%%%%%%%%%%%%%%%%%%%%%%%%%%%%%%%%%%%%%%%%%%%%%%%%%%%%%%%%%%%%%%%%%%%%%%%%%%%%%
In this section, we introduce the field theory that we work with, and the null quantization scheme that allows us to prove the Renyi QNEC. We also give the precise statement that we will prove, and reformulate it in a much more convenient form as a calculation regarding a state perturbatively close to the vacuum.

\subsection{Null quantization and the state}

\begin{figure*}[t]      \label{figure}
  \includegraphics[scale=.3]{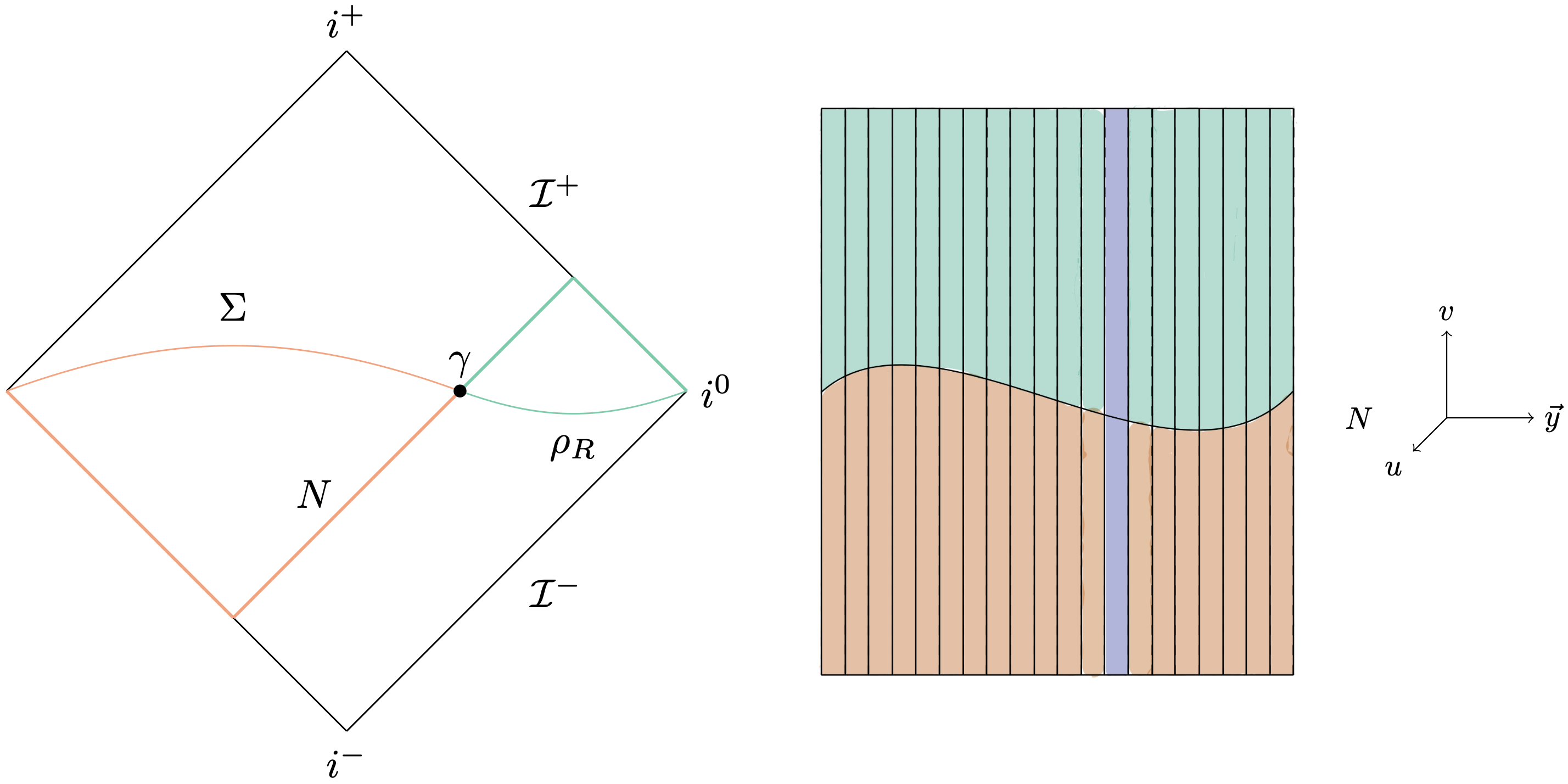}
  \caption{An illustration of the setup that we work with. $\Sigma$ is a Cauchy surface in Minkowski space, which is split in half by the codimension-2 surface $\gamma$. The reduced state, $\rho_R$, on the green half of the Cauchy surface is equivalent to the state on (green) parts of the null hypersurface $N$ and null infinity $\mathcal I^+$. The right figure is a front view of $N$. The vertical segments bounded by solid lines are the pencils on $N$, with the blue pencil being our specific pencil $P$; the rest are all part of the auxiliary system. The parameter $\l$ is $-\I$ at the bottom of $N$ in the right figure, and $\l\to\I$ at the top. The curve that divides the green and orange regions on $N$ is $\gamma$. The point $p$ used above is the intersection of $P$ with $\gamma$. The green part of $\S$ extends behind the plane of the page.}
\end{figure*}

For our proof, we will need to assume that the QFT describing the matter has a valid null-hypersurface initial-value formulation, i.e., a field algebra $\cA(N)$ can be defined on a stationary null-surface $N$ without making reference to anything outside $N.$ Our treatment of null quantization closely follows \cite{Bousso:2015wca}; see  \cite{Wall:2011hj} and \cite{Burkardt:1995ct} for more details. The details of our setup are summarized in Fig. \ref{figure}. We spell them out in detaill below.

Let $\g$ be a spacelike codimension-2 surface that splits a Cauchy surface $\S$ into two sides. The proof we present here applies when $\g$ is a section of a general stationary null surface $N$ in $D>2.$ Specifying the state of the QFT on the Cauchy surface $\S$ is unitarily equivalent to specifying the state on the null plane $N$ and parts of past and future null infinities. 

Now discretize $N$ along the transverse directions into small regions of transverse area $\sA.$ These regions fully extended along the null directions are called pencils. We use $\sA$ as an expansion parameter and take the limit $\sA\to0$ at the end. Degrees of freedom on different horizon generators are independent systems \cite{Wall:2011hj}, and so the Hilbert space on $N$ factorizes into a product of Hilbert spaces on the pencils. On each pencil, there exists a $1+1$-dimensional free chiral fermionic conformal field theory. Details of the CFT on the pencils are provided in Appendix \ref{app_fermi_cft}.

We want to consider the second shape variation along $N$ of SRD at some point $q$ on $\g$. This point is contained in one specific pencil, which we denote by $P.$ Decompose the Hilbert space of the system as $\cH=\cH_{P}\otimes\cH_{A}$, where $\cH_P$ is the Hilbert space of our specific pencil, $P,$ and $\cH_A\equiv\cH_{\rm auxiliary}$ contains all the remaining degrees of freedom. Consider a density matrix on $\cH$, which we deform to obtain a one-parameter family of density matrices $\r(\l)$ by tracing out the part of the pencil $P$ in the past of affine parameter $\l.$ For small $\sA$, the state of the pencil is near the vacuum, and the global state can be written as
\begin{equation}			\label{eq_red_den_mat}
	\r(\l) = \s_P(\l)\otimes\r_{A}^{(0)}+\sA^{1/2}\r^{(1)}(\l) + \cdots.
\end{equation}
This can be seen as follows. Since the probability to have $n$ particles on a pencil should scale extensively with the area, it is proportional to $\sA^n$ for each pencil. Then, terms of the form $\ket n\!\!\bra m$ in the pencil Fock basis scale as $\sA^{(n+m)/2}$. Thus the leading contribution to the perturbation, $\r^{(1)}(\l)$, is of order $\sA^{1/2}$, with terms of the form $\ket0\!\!\bra1$ and $\ket1\!\!\bra0$. Explicitly, the state on $N$ must take the form
\begin{align}
	\r = \ketbra{0}{0}\otimes\r_A^{(0)} +&\, \sA^{1/2}\sum_{ij}\big(\ketbra{0}{\psi_{ij}^1}+\ketbra{\psi_{ij}^2}{0}\big)\otimes\ketbra{i}{j}	\\
		 +&\, \sA^{1/2}\sum_{ij}\big(\ketbra{0}{\psi_{ij}^2}+\ketbra{\psi_{ij}^1}{0}\big)\otimes\ketbra{j}{i} + \cdots ,					\nn
\end{align}
where $\ket{\psi_{ij}}$ are single particle states on the pencil CFT which can be constructed by inserting single operators in the fermionic path integral; and $\lbr\ket i,\ket j \rbr$ together form an arbitrary basis of $\cH_A$ such that $\lbr\ketbra ij,\ketbra ji\rbr$ form a Grasssmann-odd basis of operators on $\cH_A$. The basis elements need to be Grassmann-odd since the pencil CFT has one fermion excited in the leading contribution to the perturbation. We also note for later use that physical observables can only be Grassmann-even. Note that this can be achieved, e.g., by taking $\ket i (\ket j) $ to be states with an even (odd) number of creation operators acting on the vacuum. We stick to this convention throughout the paper. A resolution of the identity operator for the auxiliary system in this basis is 
\begin{equation}
	\mathbf 1_a = \sum_{i}\ketbra ii+\sum_j\ketbra jj \equiv \sum_\mu\ketbra\mu\mu,
\end{equation}
where the sum over $i,$ respectively $j,$ runs over states with even, respectively odd, numbers of creation operators. We use the convention that indices $\mu,\nu$ can correspond to either even ($i$) or odd ($j$) states. Note that the auxiliary system, $\r_{A}^{(0)},$ is not necessarily in the vacuum state. Entanglement between the pencil and the auxiliary system is contained in $\r^{(1)}(\l)$ and the further subleading terms.

Since the theory on each pencil is chiral, we can replace translations (and derivatives) along $\l$ by translations (and derivatives) along the spatial direction. Also, the single particle states, $\ket{\psi_{ij}}$, of the Fock space can be constructed by a Euclidean path integral over the lower half plane, $\t<0,$ with insertions of a single field. 

The affine parameter $\l=V_P$, $V_P$ some constant, describes the location of the entangling surface on the pencil or, equivalently, the location $x=\l$ on the $\t=0$ slice of the Euclidean path integral. Using null translation invariance of the free theory, we choose to fix the entangling surface to be at $x=0$ by moving the operator insertions simultaneously. The reduced density matrix $\r(\l)$ then corresponds to tracing out the region $x<0.$ Deformations of the entangling region now amount to changing the location of the operator insertions, allowing us to evaluate shape derivatives with relative ease. Further, the density matrix $\s_P^{(0)}$ is now independent of $\l.$ In fact,
\begin{equation}
	\s_P = e^{-2\pi K_P},
\end{equation}
where $K_P$ is the modular Hamiltonian on the pencil \cite{Unruh:1976db,Bisognano:1976za}.

We can now write an arbitrary state as
\begin{align}
	\r(\l) = \s_P\otimes\r_{A}^{(0)}\,+&\,\sA^{1/2}\sum_{ij}\lb \s_P \int drd\q\,f_{ij}(r,\q)\psi(re^{i\q}-\l) \rb\otimes\ketbra ij	\nn	\\
	+&\, \sA^{1/2}\sum_{ij}\lb \s_P \int drd\q\,f_{ji}(r,\q)\psi(re^{i\q}-\l) \rb\otimes\ketbra ji + \cO(\sA),
\end{align}
which we write as
\begin{align}
	\r(\l) = \s_P\otimes\r_{A}^{(0)}\,+&\,\sA^{1/2}\sum_{\mn}\lb \s_P \int drd\q\,f_{\mn}(r,\q)\psi(re^{i\q}-\l) \rb\otimes\ketbra \mu\nu +  \cO(\sA).
\end{align}
Defining
\begin{align}
	\r^{(1)}(\l) =&\ \sum_{\mn}\lb \s_P \int drd\q\,f_{\mn}(r,\q)\psi(re^{i\q}-\l) \rb\otimes\ketbra \mu\nu,	\nn
\end{align}
we succinctly write
\begin{equation}			\label{eq_red_den_mat_2}
	\r(\l) = \s_P\otimes\r_{A}^{(0)} + \sA^{1/2}\r^{(1)}(\l) +  \cO(\sA),
\end{equation}
with the factor of $\sA^{1/2}$ explicit. To ensure that the state is Hermitian, we need to impose
\begin{equation}
	f_{\mn}(r,\q) = -if_{\nu\mu}^*(r,2\pi-\q),
\end{equation}
where we emphasize that $\mu,\nu$ indices can be either $i$ or $j$. 

It is necessary to require that $f_{\mn}(r,\q)$ vanishes at $\q=0,\pi$ so that the state is normalizable. To ensure that SRD is finite for the states we study, it is further necessary to restrict the support of the functions $f_{\mn}$ to a wedge of size $2\pi/n$ centered at $\q=\pi$, as discussed in Section \ref{subsec:corrfns}. We will refer to this requirement as the wedge condition.

%%%%%%%%%%%%%%%%%%%%%%%%%%%%%%%%%%%%%%%%%%%%%%%%%%%%%%%%%%%%%%%%%%%%%%%%%%%%%%%%%%%%%
%%%%%%%%%%%%%%%%%%%%%%%%%%%%%%%%%%%%%%%%%%%%%%%%%%%%%%%%%%%%%%%%%%%%%%%%%%%%%%%%%%%%%
\subsection{Statement of Renyi QNEC}
Recall that for density matrices, sandwiched Renyi divergence (SRD) of  $\r$ with respect to $\s$ is given by
\begin{equation}
	S_n^R(\r|\s) = \frac1{n-1}\log\whZ_n(\r|\s),
\end{equation}
where
\begin{equation}			\label{eq_whZn_defn}
	\whZ_n(\r|\s) = \Tr\lB\lb \s^{\frac{1-n}{2n}}\r\s^{\frac{1-n}{2n}} \rb^n \rB.
\end{equation}
For the Renyi QNEC, $\r$ is the density matrix $\r(\l)$ introduced above, defined for the subregion $v>V(\vec y)$, along the null surface $N$, and $\s$ is the vacuum density matrix. The deformation relevant for Renyi QNEC is then simply translation along the affine parameter $\l$ on the pencil $P$ near $\l=0$, and Renyi QNEC is the statement that, for $1<n\in\R$,
\begin{equation}			\label{eq_Renyi_QNEC_2}
	\left.\lim_{\sA\to0}\frac1{\sA}\frac{d^2}{d\l^2}S_n(\r(\l)|\s)\right\vert_{\l=0}\geq0.
\end{equation}

If we expand the state $\r(\l)$ as
\begin{equation}
	\r(\l) = \s_P\otimes\r_A^{(0)} + \sA^{1/2}\r^{(1)}(\l) + \sA\r^{(2)}(\l) + \cO(\sA^{3/2}),
\end{equation}
$\whZ_n$ can be expanded as
\begin{equation}
	\whZ_n(\r(\l)|\s) = \whZ_n^{(0)} + \sA^{1/2}\whZ_n^{(1)}(\l) + \sA\lb \whZ_n^{(2)}(\l) + \whZ_n^{(1,1)}(\l) \rb + \cO(\sA^{3/2}),
\end{equation}
where $\whZ_n^{(i)}$ contains contributions from the $i^{\rm th}$ term of the $\r(\l)$ expansion, and $\whZ_n^{(1,1)}$ contains two powers of $\r^{(1)}$. Using this is (\ref{eq_Renyi_QNEC_2}), one can recast the Renyi QNEC statement as \cite{Moosa:2020jwt}
\begin{equation}			\label{srd_zn11}
	\ld\frac{d^2}{d\l^2} Z_n^{(1,1)}(\l)\right\vert_{\l=0}\geq 0,
\end{equation}
where
\begin{equation}
	Z_n^{(1,1)}(\l) = \frac1{n-1}\whZ_n^{(1,1)}(\l).
\end{equation}

%%%%%%%%%%%%%%%%%%%%%%%%%%%%%%%%%%%%%%%%%%%%%%%%%%%%%%%%%%%%%%%%%%%%%%%%%%%%%%%%%%%%%
%%%%%%%%%%%%%%%%%%%%%%%%%%%%%%%%%%%%%%%%%%%%%%%%%%%%%%%%%%%%%%%%%%%%%%%%%%%%%%%%%%%%%
\subsection{Reformulation as a perturbative calculation}

As it stands, we need to perform a calculation of SRD between an arbitrary state and the vacuum. As demonstrated in \cite{Moosa:2020jwt} however, we can improve the situation and restate our problem as a calculation of SRD between two perturbatively close states. 

Let us first define
\begin{equation}
	\tilde\r^{(0)} = \s_P\otimes\tilde\r_{A}^{(0)}, \qquad \tilde\r_{A}^{(0)} = \lb \s_{a}^{\frac{1-n}{2n}}\r_{A}^{(0)}\s_{a}^{\frac{1-n}{2n}} \rb^n,
\end{equation}
and choose a basis, $\lbr\ket i,\ket j \rbr$, such that $\tilde\r_A^{(0)}$ acts diagonally,
\begin{equation}
	\tilde\r_A^{(0)} \ket\mu = e^{-2\pi K_\mu}\ket\mu.
\end{equation}
We also define  
\begin{equation}
	E_{\mn}\equiv e^{\q(K_\mu-K_\nu)}\ket\mu\!\!\bra\nu.
\end{equation}
Then, inserting a complete set of states, we can write
\begin{align}
	\s^{\frac{1-n}{2n}}\r^{(1)}(\l)\s^{\frac{1-n}{2n}} =	\sum_{\mu_1\mn\mu_2}&\,\s_P^{\frac{1-n}{2n}+1}\lb\int drd\q\,f_{\mn}(r,\q)\psi(re^{i\q}-\l)\rb\s_P^{\frac{1-n}{2n}}	\nn	\\
	&\otimes \ketbra{\mu_1}{\mu_1} \s_a^{\frac{1-n}{2n}}\ketbra \mu\nu\s_a^{\frac{1-n}{2n}}\ketbra {\mu_2}{\mu_2}.
\end{align}
Note that since $\s_a$ is a Grassmann-even operator, elements such as $\bra i\s_a^m\ket j$ will vanish.

Now define
\begin{equation}
	\tilde f_{\mn}(r,\q) = \sum_{\mu_1\nu_1}f_{\mu_1\nu_1}(r,\q)\bra\mu\s_a^{\frac{1-n}{2n}}\ketbra{\mu_1}{\nu_1}\s_a^{\frac{1-n}{2n}}\ket\nu,
\end{equation}
%Note that $\tilde f_{\mn}$ 
which satisfies the same reality conditions as $f_{\mn}$. This allows us to write
\begin{align}
	\s^{\frac{1-n}{2n}}\r^{(1)}(\l)\s^{\frac{1-n}{2n}} =&\ \sum_{\mn} \s_P^{\frac{1-n}{2n}}\lb  \s_P\int drd\q\tilde f_{\mn}(r,\q)\psi(re^{i\q}-\l) \rb\s_P^{\frac{1-n}{2n}}\otimes\ket\mu\!\!\bra\nu.
\end{align}
Note that any function that satisfies the same reality and support conditions as $f_{\mn}(r,\q)$ is equally valid as a support function for the field. We use this freedom to write \cite{Moosa:2020jwt}
\begin{align}
	\s^{\frac{1-n}{2n}}\r^{(1)}(\l)\s^{\frac{1-n}{2n}} =&\ \sum_{\mn} \s_P^{\frac{1-n}{2n}}\lb  \s_P\int drd\q\tilde f_{\mn}(r,\q)\psi(re^{i\q}-\l) \rb\s_P^{\frac{1-n}{2n}}		\nn		\\
	& \qquad\qquad \otimes \lb (\tilde\r_A^{(0)})^{\frac{1-n}{2n}}\tilde\r_A^{(0)}E_{\mn}(\q)(\tilde\r_A^{(0)})^{\frac{1-n}{2n}} \rb.
\end{align}
Defining
\begin{equation}			\label{eq_tilde_rho1_defn}
	\tilde\r^{(1)}(\l) = \sum_{\mn}\tilde\r^{(0)}\int drd\q \tilde f_{\mn}(r,\q)\lb \psi(re^{i\q}-\l)\otimes E_{\mn}(\q) \rb,
\end{equation}
we have the result that
\begin{align}
	Z_n(\l) =&\ \frac1{n-1}\Tr\lB\lb \s^{\frac{1-n}{2n}}\r\s^{\frac{1-n}{2n}} \rb^n \rB		\nn		\\
	 =&\ \frac1{n-1}\Tr\lB \lb \big(\tilde\r^{(0)}\big)^{1/n} + \sA^{1/2}\big(\tilde\r^{(0)}\big)^{\frac{1-n}{2n}}\tilde\r^{(1)}(\l)\big(\tilde\r^{(0)}\big)^{\frac{1-n}{2n}} \rb^n\rB.
\end{align}
This is equivalent to
\begin{equation}			\label{eq_pert_RQNEC}
	Z_n(\l) = Z_n\left(\tilde\r^{(0)}+\sA^{1/2}\tilde\r^{(1)}(\l)|\tilde\r^{(0)}\right),
\end{equation}
which is just the SRD between two perturbatively close states. This quantity has been studied previously in \cite{May:2018tir,Moosa:2020jwt}, and is much simpler than the original statement (\ref{srd_zn11}) involving a state arbitrarily far from the vacuum.

%%%%%%%%%%%%%%%%%%%%%%%%%%%%%%%%%%%%%%%%%%%%%%%%%%%%%%%%%%%%%%%%%%%%%%%%%%%%%%%%%%%%%
%%%%%%%%%%%%%%%%%%%%%%%%%%%%%%%%%%%%%%%%%%%%%%%%%%%%%%%%%%%%%%%%%%%%%%%%%%%%%%%%%%%%%
%%%%%%%%%%%%%%%%%%%%%%%%%%%%%%%%%%%%%%%%%%%%%%%%%%%%%%%%%%%%%%%%%%%%%%%%%%%%%%%%%%%%%
\section{Proof}						\label{sec:proof}

We prove Renyi QNEC in two ways. We first deal with the case where the Renyi index $1<n\in\Z^+$. This proof is much shorter and neater, following just from reflection positivity, the Euclidean analogue of unitarity. Then we proceed to calculate the second variation of SRD for arbitrary $n$, and calculate correlation functions of the pencil and the auxiliary systems to show that the relevant shape variation is in fact positive.

%%%%%%%%%%%%%%%%%%%%%%%%%%%%%%%%%%%%%%%%%%%%%%%%%%%%%%%%%%%%%%%%%%%%%%%%%%%%%%%%%%%%%
%%%%%%%%%%%%%%%%%%%%%%%%%%%%%%%%%%%%%%%%%%%%%%%%%%%%%%%%%%%%%%%%%%%%%%%%%%%%%%%%%%%%%
\subsection{For integer $n>1$}

For $n\in\Z^+,$ it is easy to see from (\ref{eq_pert_RQNEC}) that
\begin{equation}			\label{eq_Zn11_sumk}
	Z_n^{(1,1)}(\l) = \frac n{2(n-1)}\sum_{k=1}^{n-1}\Tr\lb (\tilde\r^{(0)})^{-1+\frac kn}\tilde\r^{(1)}(\l)(\tilde\r^{(0)})^{-\frac kn}\tilde\r^{(1)}(\l) \rb.
\end{equation}
Denoting $\l$-derivatives by dots, we are interested in the second derivative
\begin{align}
	\ddot Z_n^{(1,1)}(\l)	= \frac n{2(n-1)}\sum_{k=1}^{n-1}\bigg[&\Tr \bigg((\tilde\r^{(0)})^{-1+\frac kn} \ddot{\tilde\r}^{(1)}(\l)(\tilde\r^{(0)})^{-\frac kn}\tilde\r^{(1)}(\l) \bigg) \nn \\
			&  + 2\,\Tr\lb(\tilde\r^{(0)})^{-1+\frac kn}\dot{\tilde\r}^{(1)}(\l)(\tilde\r^{(0)})^{-\frac kn}\dot{\tilde\r}^{(1)}(\l)\rb   \\
			& + \Tr\lb(\tilde\r^{(0)})^{-1+\frac kn}\tilde\r^{(1)}(\l)(\tilde\r^{(0)})^{-\frac kn}\ddot{\tilde\r}^{(1)}(\l) \rb\bigg].	\nn
\end{align}
Using the explicit form of $\tilde\r^{(1)}$ given in (\ref{eq_tilde_rho1_defn}) and setting $\l=0$, this becomes
\begin{align}
	\ddot Z_n^{(1,1)}	= \frac n{2(n-1)}\sum_{k=1}^{n-1}\int d\a\bigg[&\Tr\lb \tilde\r^{(0)} (\tilde\r^{(0)})^{-\frac kn} \cO_{\mu_2\nu_2}(r_2,\q_2)  (\tilde\r^{(0)})^{\frac kn} \ddot{\cO}_{\mu_1\nu_1}(r_1,\q_1)	\rb	\nn		\\
	&\ + 2\,\Tr\lb \tilde\r^{(0)} (\tilde\r^{(0)})^{-\frac kn} \dot{\cO}_{\mu_2\nu_2}(r_2,\q_2)  (\tilde\r^{(0)})^{\frac kn} \dot{\cO}_{\mu_1\nu_1}(r_1,\q_1)\rb	\\
	&\ + \Tr\lb \tilde\r^{(0)} (\tilde\r^{(0)})^{-\frac kn} \ddot{\cO}_{\mu_2\nu_2}(r_2,\q_2)  (\tilde\r^{(0)})^{\frac kn} {\cO}_{\mu_1\nu_1}(r_1,\q_1)\rb\bigg],	\nn
\end{align}
where we have defined
\begin{equation}			\label{eq_dalpha_defn}
	\int d\a = \sum_{\mu_2\nu_2}\sum_{\mu_1\nu_1}\int dr_1dr_2\int_{\pi-\pi/n}^{\pi+\pi/n}d\q_1d\q_2\tilde f_{\mu_1\nu_1}(r_1,\q_1) \tilde f_{\mu_2\nu_2}(r_2,\q_2),
\end{equation}
and
\begin{align}				\label{eq_cO_defn}
	\cO_{\mn}	= \psi(re^{i\q})\otimes E_{\mn}(\q), \quad \dot{\cO}_{\mn}	= \del\psi(re^{i\q})\otimes E_{\mn}(\q),\quad \ddot{\cO}_{\mn}	= \del^2\psi(re^{i\q})\otimes E_{\mn}(\q).
\end{align}
Since the conformal weight of $\del^m\psi$ is $(\frac12+m,0)$, it transforms under conjugation by $\s_P$ as
\begin{equation}			\label{eq_psi_mod_flow}
	\s_P^{-\frac k{2n}}\del^m\psi(re^{i\q})\s_P^{\frac k{2n}} = e^{i\pi\frac kn(m+\frac12)}\del^m\psi(re^{i\q+i\pi\frac kn}).
\end{equation}
The other factor, $E_{\mn}(\q)$, tranforms under conjugation by $\tilde\r_A^{(0)}$ as
\begin{equation}
	(\tilde\r_A^{(0)})^{-\frac k{2n}}E_{\mn}(\q)(\tilde\r_A^{(0)})^{\frac k{2n}} = E_{\mn}(\q+\pi k/n).
\end{equation}
Combining these, we can conjugate the operators $\cO_{\mn}$ and their derivatives to get
\begin{align}
	\ddot Z_n^{(1,1)}	= \frac n{2(n-1)}\sum_{k=1}^{n-1}\int d\a\bigg[& e^{-i\frac{5\pi k}{2n}}\Tr\lb\tilde\r^{(0)} \cO_{\mu_2\nu_2}(r_2,\q_2+\pi k/n)\ddot\cO_{\mu_1\nu_1}(r_1,\q_1-\pi k/n)\rb \nn \\
		+\, 2&\, \Tr\lb\tilde\r^{(0)} \dot\cO_{\mu_2\nu_2}(r_2,\q_2+\pi k/n)\dot\cO_{\mu_1\nu_1}(r_1,\q_1-\pi k/n)\rb \\
		+&\, e^{i\frac{5\pi k}{2n}}\Tr\lb\tilde\r^{(0)} \ddot\cO_{\mu_2\nu_2}(r_2,\q_2+\pi k/n)\cO_{\mu_1\nu_1}(r_1,\q_1-\pi k/n) \rb\bigg].\nn
\end{align}
Due to the wedge condition, the opertor insertions are angle-ordered and the terms above can be written as correlation functions,
\begin{align}
	\ddot Z_n^{(1,1)}	= \frac n{2(n-1)}\sum_{k=1}^{n-1}\int d\a\bigg[& e^{-i\frac{5\pi k}{2n}}\left\langle \cO_{\mu_2\nu_2}(r_2,\q_2+\pi k/n)\ddot\cO_{\mu_1\nu_1}(r_1,\q_1-\pi k/n) \right\rangle \nn \\
		&\ + 2\left\langle \dot\cO_{\mu_2\nu_2}(r_2,\q_2+\pi k/n)\dot\cO_{\mu_1\nu_1}(r_1,\q_1-\pi k/n) \right\rangle	\\
		&\ + e^{i\frac{5\pi k}{2n}}\left\langle \ddot\cO_{\mu_2\nu_2}(r_2,\q_2+\pi k/n)\cO_{\mu_1\nu_1}(r_1,\q_1-\pi k/n) \right\rangle\bigg].	\nn
\end{align}
Now we note that since
\begin{equation}
	\braket{\del_z^2\psi(z)\psi(w)}_p = - \braket{\del_z\psi(z)\del_w\psi(w)}_p = \braket{\psi(z)\del_w^2\psi(w)}_p,
\end{equation}
the correlators appearing in $\ddot Z_n^{(1,1)}$ are in fact all proportional to each other. This allows us to write
\begin{align}
	\ddot Z_n^{(1,1)}	= \frac {2n}{n-1}\sum_{k=1}^{n-1}\int d\a \sin^2\lb \frac {5\pi k}{4n} \rb \left\langle \dot\cO_{\mu_2\nu_2}(r_2,\q_2+\pi k/n)\dot\cO_{\mu_1\nu_1}(r_1,\q_1-\pi k/n) \right\rangle.
\end{align}
Defining the (Grassmann-even) operators
\begin{equation}
	\begin{split}
		\Psi_k	=&\	\sum_{\mn}\int dr\int_{\pi-\pi/n}^{\pi+\pi/n}d\q\tilde f_{\mn}(r,\q)\dot\cO_{\mu\nu}(r,\q-\pi k/n),		\\
		\overbar\Psi_k	=&\ \sum_{\mn}\int dr\int_{\pi-\pi/n}^{\pi+\pi/n}d\q\tilde f_{\mn}(r,\q)\dot\cO_{\mu\nu}(r,\q+\pi k/n),
	\end{split}
\end{equation}
we see that
\begin{equation}			\label{eq_znddot_int_n}
	\ddot Z_n^{(1,1)}	= \frac {2n}{n-1}\sum_{k=1}^{n-1}\sin^2\lb \frac {5\pi k}{4n} \rb\braket{\overline\Psi_k\Psi_k}.
\end{equation}
Under the change of variables $\q\to2\pi-\q$, we have $f_{\mn}(r,2\pi-\q)=-if_{\nu\mu}^*(r,\q)$ and
\begin{equation}
	\dot\cO_{\mn}(r,2\pi-\q) = i(\tilde\r^{(0)})^{-1}\dot\cO_{\nu\mu}^\dagger(r,\q)\tilde\r^{(0)}.
\end{equation}
Using both of these, we get
\begin{equation}
	\overline\Psi_k = (\tilde\r^{(0)})^{-1}\Psi_k^\dagger\tilde\r^{(0)}.
\end{equation}
The correlators in $\ddot Z_n^{(1,1)}$ now become
\begin{equation}
	\braket{\overline\Psi_k\Psi_k} = \braket{(\tilde\r^{(0)})^{-1}\Psi_k^\dagger\tilde\r^{(0)}\Psi_k} = \braket{\Psi_k\Psi_k^\dagger}>0,
\end{equation}
where the last inequality is just the statement of reflection positivity\footnote{See, e.g., \cite{Simmons-Duffin:2016gjk}, for an overview of reflection positivity.}. Thus all terms in (\ref{eq_znddot_int_n}) are individually positive, and we have proved Renyi QNEC for integer $n>1.$

%%%%%%%%%%%%%%%%%%%%%%%%%%%%%%%%%%%%%%%%%%%%%%%%%%%%%%%%%%%%%%%%%%%%%%%%%%%%%%%%%%%%%
%%%%%%%%%%%%%%%%%%%%%%%%%%%%%%%%%%%%%%%%%%%%%%%%%%%%%%%%%%%%%%%%%%%%%%%%%%%%%%%%%%%%%
\subsection{Second variation of SRD for arbitrary $n$}

Let $\ket\k$ denote a basis in which $\tilde\r^{(0)}$ is diagonal, i.e.,
\begin{equation}
	\tilde\r^{(0)}\ket\k = e^{-2\pi\k}\ket\k,
\end{equation}
where $\k$ can be negative since $\tilde\r^{(0)}$ is not a normalized density matrix. Recall from (\ref{eq_Zn11_sumk}) that
\begin{equation}
	Z_n^{(1,1)}(\l) = \frac n{2(n-1)}\sum_{k=1}^{n-1}\Tr\lb (\tilde\r^{(0)})^{-1+\frac kn}\tilde\r^{(1)}(\l)(\tilde\r^{(0)})^{-\frac kn}\tilde\r^{(1)}(\l) \rb.
\end{equation}

One can show that, in this basis, we have the following result for arbitrary $n$ \cite{Moosa:2020jwt}
\begin{equation}
	Z_{n}^{(1,1)}(\l) = \frac12\int d\k\int d\k' e^{2\pi\k'}F_n(\k-\k')\left\vert\left\langle\k\left\vert\tilde\r^{(1)}(\l)\right\vert\k'\right\rangle\right\vert^2,
\end{equation}
where
\begin{equation}
	F_n(x) = -\frac n{n-1}\frac{e^{2\pi\lb \frac{n-1}n\rb x}-1}{e^{-2\pi x/n}-1}.
\end{equation}

Taking two derivatives with respect to $\l$ gives us
\begin{align}
	\ddot Z_n^{(1,1)}(\l) =&\ \int d\k d\k' e^{2\pi\k'} F_n(\k-\k')	\\	
		&\ \qquad \times\lb \langle\k\vert\tilde\r^{(1)}(\l)\vert\k'\rangle\langle\k'\vert\ddot{\tilde\r}^{(1)}(\l)\vert\k\rangle + \langle\k\vert\dot{\tilde\r}^{(1)}(\l)\vert\k'\rangle\langle\k'\vert\dot{\tilde\r}^{(1)}(\l)\vert\k\rangle \rb	,	\nn
\end{align}
where dots again denote $\l$-derivatives. Using the definition of $\tilde\r^{(1)}$ in (\ref{eq_tilde_rho1_defn}) and setting $\l=0$,
\begin{align}
	\ld\ddot Z_n^{(1,1)}\rv_{\l=0} &= \int d\a\int d\k d\k' e^{-2\pi\k}F_n(\k-\k')		\\
		 \times\Big(& \langle\k\vert\cO_{\mu_1\nu_1}(r_1,\q_1)\vert\k'\rangle\langle\k'\vert\ddot\cO_{\mu_2\nu_2}(r_2,\q_2)\vert\k\rangle + \langle\k\vert\dot\cO_{\mu_1\nu_1}(r_1,\q_1)\vert\k'\rangle\langle\k'\vert\dot\cO_{\mu_2\nu_2}(r_2,\q_2)\vert\k\rangle \Big),		\nn
\end{align}
where $\int d\a$ is defined in (\ref{eq_dalpha_defn}) and $\cO_{\mn}$ in (\ref{eq_cO_defn}).

Making the angular-ordering explicit, this is
\begin{align}
	&\ddot Z_n^{(1,1)}	=	\int_{\q_1>\q_2}d\a\int d\k d\k'e^{-2\pi\k}F_n(\k-\k')		\\
		&\qquad \times \lb \langle\k\vert\cO_{\mu_1\nu_1}(r_1,\q_1)\vert\k'\rangle\langle\k'\vert\ddot\cO_{\mu_2\nu_2}(r_2,\q_2)\vert\k\rangle + \langle\k\vert\dot\cO_{\mu_1\nu_1}(r_1,\q_1)\vert\k'\rangle\langle\k'\vert\dot\cO_{\mu_2\nu_2}(r_2,\q_2)\vert\k\rangle \rb		\nn		\\
		 &+ \int_{\q_2>\q_1}d\a\int d\k d\k'e^{-2\pi\k'}F_n(\k'-\k)	\nn		\\
		&\qquad \times \lb \langle\k\vert\cO_{\mu_1\nu_1}(r_1,\q_1)\vert\k'\rangle\langle\k'\vert\ddot\cO_{\mu_2\nu_2}(r_2,\q_2)\vert\k\rangle + \langle\k\vert\dot\cO_{\mu_1\nu_1}(r_1,\q_1)\vert\k'\rangle\langle\k'\vert\dot\cO_{\mu_2\nu_2}(r_2,\q_2)\vert\k\rangle \rb,		\nn	
\end{align}
where we have used the fact that $F_n(x)=e^{2\pi x}F_n(-x).$ Decomposing $F_n(x)$ into Fourier modes as
\begin{equation}
	F_n(x) = \int_{-\I}^\I ds\,e^{isx}\cF_n(s),
\end{equation}
one obtains
\begin{align}
	\ddot Z_n^{(1,1)}	=&\ \int_{\q_1>\q_2}d\a\int_{-\I}^\I ds\cF_n(s) \bigg\{ \Tr\lB (\tilde\r^{(0)})^{1-\frac{is}{2\pi}}\cO_{\mu_1\nu_1}(r_1,\q_1)(\tilde\r^{(0)})^{\frac{is}{2\pi}}\ddot\cO_{\mu_2\nu_2}(r_2,\q_2)\rB		\\
		&\ \qquad\qquad\qquad\qquad\qquad  +\Tr \lB (\tilde\r^{(0)})^{1-\frac{is}{2\pi}}\dot\cO_{\mu_1\nu_1}(r_1,\q_1)(\tilde\r^{(0)})^{\frac{is}{2\pi}}\dot\cO_{\mu_2\nu_2}(r_2,\q_2)\rB\bigg\}	\nn		\\
		&\ +\int_{\q_2>\q_1}d\a\int_{-\I}^\I ds\cF_n(-s) \bigg\{ \Tr\lB (\tilde\r^{(0)})^{1+\frac{is}{2\pi}}\ddot\cO_{\mu_2\nu_2}(r_2,\q_2)(\tilde\r^{(0)})^{-\frac{is}{2\pi}}\cO_{\mu_1\nu_1}(r_1,\q_1) \rB	\nn		\\
		&\ \qquad\qquad\qquad\qquad\qquad  +\Tr \lB (\tilde\r^{(0)})^{1+\frac{is}{2\pi}}\dot\cO_{\mu_2\nu_2}(r_2,\q_2) (\tilde\r^{(0)})^{-\frac{is}{2\pi}}\dot\cO_{\mu_1\nu_1}(r_1,\q_1) \rB\bigg\}.	\nn
\end{align}
Since correlation functions are defined to be implicitly ordered in angular time, we can now write
\begin{equation}
	\ddot Z_n^{(1,1)} = \int d\a\int_{-\I}^\I ds\cF_n({\rm sign}(\q_{12})s)\cG(s),
\end{equation}
where we have defined $\q_{ij}=\q_i-\q_j$ and
\begin{align}
	\cG(s) =&\ \la (\tilde\r^{(0)})^{-\frac{is}{2\pi}}\cO_{\mu_1\nu_1}(r_1,\q_1)(\tilde\r^{(0)})^{\frac{is}{2\pi}}\ddot\cO_{\mu_2\nu_2}(r_2,\q_2) \ra	\\
		 &\ \qquad + \la (\tilde\r^{(0)})^{-\frac{is}{2\pi}}\dot\cO_{\mu_1\nu_1}(r_1,\q_1)(\tilde\r^{(0)})^{\frac{is}{2\pi}}\dot\cO_{\mu_2\nu_2}(r_2,\q_2) \ra.	\nn
\end{align}
Using the Fourier transform of $\cG(s)$, defined by
\begin{equation}
	G(\w) = \frac1{2\pi}\int_{-\I}^\I ds\,e^{-is\w}\cG(s),
\end{equation}
we obtain
\begin{equation}
	\ddot Z_n^{(1,1)} = \int d\a\int_{-\I}^\I d\w\,F_n({\rm sign}(\q_{12})\w)G(\w).
\end{equation}
We write this instead as
\begin{equation}
	\ddot Z_n^{(1,1)} = \int d\a\int_{-\I}^\I d\w\widetilde F_n(\w)e^{{\rm sign}(\q_{12})\pi\w}G(\w),
\end{equation}
with
\begin{equation}
	\widetilde F_n(\w) = e^{-\pi\w}F_n(\w) = \frac n{n-1}\frac{\sinh\pi\w\frac{n-1}n}{\sinh\pi\w/n}
\end{equation}
The correlation function $\cG(s)$ and its Fourier transform $G(\w)$ are calculated explicitly in Appendix \ref{app_corr_fns}. The result we obtain is
\begin{align}
	G(\w)	=&\ -\frac i8 \d_{\mu_1\nu_2}\d_{\mu_2\nu_1} e^{-\pi (K_{\mu_1}+K_{\mu_2})} e^{-{\rm sign}(\q_{12})\pi\w} (r_1e^{i\q_1})^{-\frac32}(r_2e^{i\q_2})^{-\frac32} \lb\frac{r_1}{r_2}\rb^{iK_{\mu_1\mu_2}}	\nn	\\
	&\ \quad \times \lB Q(z-i)\lb\frac{r_1e^{i\q_1}}{r_2e^{i\q_2}}\rb^{1-i\w} + Q(z)\lb\frac{r_1e^{i\q_1}}{r_2e^{i\q_2}}\rb^{-i\w}\rB,	
\end{align}
where $z=K_{\mu_1\mu_2}-\w$ and 
\begin{equation}
	Q(x) = \frac{4x^2+1}{\cosh\pi x}.
\end{equation}
We finally have that
\begin{align}
	\ddot Z_n^{1,1} =&\ -\frac i8\int d\tilde\a\, e^{-\pi (K_{\mu}+K_\nu)} (r_1e^{i\q_1})^{-\frac32}(r_2e^{i\q_2})^{-\frac32} \lb\frac{r_1}{r_2}\rb^{iK_{\mn}} \\
	&\ \times \int_{-\I}^\I d\w\, \widetilde F_n(\w) \lB Q(z)\lb\frac{r_1e^{i\q_1}}{r_2e^{i\q_2}}\rb^{-i\w} +  Q(z-i)\lb\frac{r_1e^{i\q_1}}{r_2e^{i\q_2}}\rb^{1-i\w} \rB, \nn 
\end{align}
where
\begin{equation}
	\int d\tilde\a = \sum_{\mn}\int dr_1dr_2\int d\q_1d\q_2\tilde f_{\mn}(r_1,\q_1)\tilde f_{\nu\mu}(r_2,\q_2).
\end{equation}

%%%%%%%%%%%%%%%%%%%%%%%%%%%%%%%%%%%%%%%%%%%%%%%%%%%%%%%%%%%%%%%%%%%%%%%%%%%%%%%%%%%%%
%%%%%%%%%%%%%%%%%%%%%%%%%%%%%%%%%%%%%%%%%%%%%%%%%%%%%%%%%%%%%%%%%%%%%%%%%%%%%%%%%%%%%
\subsection{Proving Renyi QNEC for $n>1$}

Note that $Q(K_{\mn}-\w)$ has no poles in the strip $-1/2\leq{\rm Im}(\w)\leq 1/2,$ and that the poles of $\widetilde F_n(\w)$ are at $\w=inp,p\in\Z$. Then, for $n>1$, we can make a contour deformation $\w\to\w\pm i/2$ without crossing any poles. This leads to the equalities
\begin{align}
  \int_{-\I}^\I d\w\, \widetilde F_n(\w)Q(z) \lb \frac{r_1e^{i\q_1}}{r_2e^{i\q_2}} \rb^{-i\w} =&\ \int_{-\I}^\I d\w\, \widetilde F_n(\w-i/2)Q(z+i/2)\lb \frac{r_1e^{i\q_1}}{r_2e^{i\q_2}} \rb^{-i\w-1/2},	\\
  \int_{-\I}^\I d\w\, \widetilde F_n(\w)Q(z-i) \lb \frac{r_1e^{i\q_1}}{r_2e^{i\q_2}} \rb^{1-i\w} =&\ \int_{-\I}^\I d\w\, \widetilde F_n(\w-i/2)Q(z-i/2)\lb \frac{r_1e^{i\q_1}}{r_2e^{i\q_2}} \rb^{-i\w+1/2}. \nn
\end{align}
We note that this deformation makes use of the fact that the source functions $\tilde f_{\mn}$ are non-vanishing only inside $|\q-\pi|<\pi/n$, in which range the integrand vanishes as ${\rm Re}(\w)\to\pm\I.$ Using these relations, we can write
\begin{align}
  \ddot Z_n^{1,1} =&\ -\frac i8\int d\tilde\a\, e^{-\pi (K_{\mu}+K_\nu)} (r_1e^{i\q_1})^{-\frac32}(r_2e^{i\q_2})^{-\frac32} \lb\frac{r_1}{r_2}\rb^{iK_{\mn}} \\
  &\ \times \int_{-\I}^\I d\w \widetilde F_n(\w-i/2)\lB Q(z+i/2)\lb\frac{r_1e^{i\q_1}}{r_2e^{i\q_2}} \rb^{-i\w-1/2} + Q(z-i/2)\lb\frac{r_1e^{i\q_1}}{r_2e^{i\q_2}} \rb^{-i\w+1/2}   \rB.   \nn
\end{align}
Under $\mu\leftrightarrow\nu,(r_1,\q_1)\leftrightarrow(r_2,\q_2),\w\to-\w,$ we get
\begin{align}
  \ddot Z_n^{1,1} =&\ -\frac i8\int d\tilde\a\, e^{-\pi (K_{\mu}+K_\nu)} (r_1e^{i\q_1})^{-\frac32}(r_2e^{i\q_2})^{-\frac32} \lb\frac{r_1}{r_2}\rb^{iK_{\mn}} \\
  &\ \times \int_{-\I}^\I d\w \widetilde F_n(\w+i/2)\lB Q(z-i/2)\lb\frac{r_1e^{i\q_1}}{r_2e^{i\q_2}} \rb^{-i\w+1/2} + Q(z+i/2)\lb\frac{r_1e^{i\q_1}}{r_2e^{i\q_2}} \rb^{-i\w-1/2}   \rB.   \nn
\end{align}
This gives us
\begin{align}
  \ddot Z_n^{1,1} =&\ -\frac i{16}\int d\tilde\a\, e^{-\pi (K_{\mu}+K_\nu)} (r_1e^{i\q_1})^{-\frac32}(r_2e^{i\q_2})^{-\frac32} \lb\frac{r_1}{r_2}\rb^{iK_{\mn}} \\
   \times \int_{-\I}^\I d\w& \lB \widetilde F_n(\w+i/2) + \widetilde F_n(\w-i/2) \rB\lB Q(z+i/2)\lb\frac{r_1e^{i\q_1}}{r_2e^{i\q_2}} \rb^{-i\w-1/2} + Q(z-i/2)\lb\frac{r_1e^{i\q_1}}{r_2e^{i\q_2}} \rb^{-i\w+1/2}   \rB.   \nn
\end{align}
We want to show that this expression is positive definite. It can be written as
\begin{align*}
  &\ -\frac i{16}\sum_{\mn}e^{-\pi(K_\mu+K_\nu)}\int_{-\I}^\I d\w \lB \widetilde F_n(\w+i/2) + \widetilde F_n(\w-i/2) \rB\   \\
  &\ \times\bigg[ Q(K_{\mn}-\w+i/2)\int dr_1dr_2\int d\q_1d\q_2\tilde f_{\mn}(r_1,\q_1)r_1^{iK_{\mn}-i\w-2}e^{-(2i-\w)\q_1}\tilde f_{\nu\mu}(r_2,\q_2)r_2^{-iK_{\mn}+i\w-1}e^{-(i+\w)\q_2}   \\
  &\ +   Q(K_{\mn}-\w-i/2)\int dr_1dr_2\int d\q_1d\q_2\tilde f_{\mn}(r_1,\q_1)r_1^{iK_{\mn}-i\w-1}e^{-(i-\w)\q_1}\tilde f_{\nu\mu}(r_2,\q_2)r_2^{-iK_{\mn}+i\w-2}e^{-(2i+\w)\q_2} \bigg]
\end{align*}
One can show that for $n\geq1,$
\[
   \widetilde F_n(\w+i/2) + \widetilde F_n(\w-i/2) = -\frac{2n}{n-1}\frac{\cosh\pi\w\sin\frac\pi n}{\cos\frac\pi n - \cosh\frac{2\pi\w}n} \leq0.
\]
We now make the substitution $\q_2\to 2\pi-\q_2 $, and use $\tilde f_{\mn}(r,\q) = -i\tilde f_{\nu\mu}^*(r,2\pi-\q)$. The term within the square brackets above then becomes
\begin{align*}
 -ie^{-2\pi\w}\big[ Q(K_{\mn}-\w+i/2)\int dr_1dr_2\int d\q_1d\q_2\tilde f_{\mn}(r_1,\q_1)r_1^{iK_{\mn}-i\w-2}e^{-(2i-\w)\q_1}\tilde f_{\mu\nu}^*(r_2,\q_2)r_2^{-iK_{\mn}+i\w-1}e^{(i+\w)\q_2}   \\
  +   Q(K_{\mn}-\w-i/2)\int dr_1dr_2\int d\q_1d\q_2\tilde f_{\mn}(r_1,\q_1)r_1^{iK_{\mn}-i\w-1}e^{-(i-\w)\q_1}\tilde f_{\mu\nu}^*(r_2,\q_2)r_2^{-iK_{\mn}+i\w-2}e^{(2i+\w)\q_2}\big]
\end{align*}

If we now again perform the substitutions $\w\to\w\pm i/2$ in the first and second terms respectively, we get
\begin{align}
  &\ddot Z_n^{1,1} = \frac n{4(n-1)}\sum_{\mn}e^{-\pi(K_\mu+K_\nu)}\int_{-\I}^\I d\w\, e^{-2\pi\w}\sin^2\lb\frac\pi n\rb\frac{\sinh\pi\w\coth\frac{\pi\w}n}{\cosh\frac{2\pi\w}n-\cos\frac{2\pi}n}  \\
  &\times Q(K_{\mn}-\w)\int dr_1d\q_1\tilde f_{\mn}(r_1\q_1)r_1^{iz-3/2}e^{(-\frac{3i}2+\w)\q_1}  \times \int dr_2d\q_2\tilde f_{\mn}^*(r_2,\q_2) r_2^{-iz-3/2}e^{(\frac{3i}2+\w)\q_2}.  \nn
\end{align}
We note that the last two integrals are complex conjugates of each other, and the other terms are all positive for $n\geq1$. Thus,
\begin{equation}
  \ddot Z_n^{1,1}\geq 0,
\end{equation}
proving the Renyi QNEC for free fermions for arbitrary $n\geq 1.$ We note that the limit $n\to 1^+$ correctly reproduces the answer in \cite{Malik:2019dpg}.

%%%%%%%%%%%%%%%%%%%%%%%%%%%%%%%%%%%%%%%%%%%%%%%%%%%%%%%%%%%%%%%%%%%%%%%%%%%%%%%%%%%%%
%%%%%%%%%%%%%%%%%%%%%%%%%%%%%%%%%%%%%%%%%%%%%%%%%%%%%%%%%%%%%%%%%%%%%%%%%%%%%%%%%%%%%
%%%%%%%%%%%%%%%%%%%%%%%%%%%%%%%%%%%%%%%%%%%%%%%%%%%%%%%%%%%%%%%%%%%%%%%%%%%%%%%%%%%%%
\section{Discussion}

Sandwiched Renyi divergence is a new measure of distance between states in Hilbert spaces, which has the desirable properties of being positive and satisfying the data processing inequality.
Previous studies of SRD in the context of QFT and holography include \cite{May:2018tir, Casini:2018cxg, Ugajin:2018rwd, Brehm:2020zri, Caginalp:2022uzd}. Motivated by the purely information theoretic formulation of QNEC, a Renyi QNEC was conjectured \cite{Lashkari:2018nsl}, and was soon proven for free bosons \cite{Moosa:2020jwt}. In this work, we have generalized the arguments of \cite{Moosa:2020jwt} to the case of free fermions, and showed that the Renyi QNEC indeed holds for $n\geq1.$

Other related distance measures between quantum states are the Petz divergence \cite{Petz:1985inf,Petz:1986fin,Lashkari:2018nsl}, the $\a$-$z$-Renyi relative entropy \cite{Audenaert:2013ere}, optimized quantum $f$-divergences \cite{Wilde:2017okz, Gao:2020gzh} and the refined Renyi divergence defined in \cite{Bao:2019aol}. A visual summary of the interrelations between various entropy measures is found in \cite{entropyzoo}. Another measure is the recently defined multi-state quantum $f$-divergence \cite{Furuya:2021kqx}. A Renyi mutual information in QFT was defined very recently in \cite{Kudler-Flam:2022zgm}. It is natural to expect that some of these measures also satisfy a QNEC like contraint on second null shape derivatives. This was already pointed out for some divergences in \cite{Moosa:2020jwt}. It will be very interesting to see if the techniques used in this work can be used to check such conjectural inequalities, and we hope to report on this front in a future work. An important technical direction to explore is proving Renyi QNEC beyond the free regime using methods of algebraic QFT as in \cite{Ceyhan:2018zfg}.

Inverting the question, it is as important to find and understand examples where such general inequalities fail to hold. The first example to demonstrate the violation of QNEC to the author's best knowledge is \cite{Ishibashi:2018ern}, which attributed the violation to IR effects. In \cite{Ecker:2019ocp}, the authors tudied the evolution of QNEC after a quench in AdS$_3$/CFT$_2$. \cite{Kibe:2021qjy} found very interestingly that QNEC can in fact be violated in these situations, and that non-violation of QNEC places bounds on the thermodynamics of the system post-quench. This was further studied in \cite{Banerjee:2022dgv} in the context of inhomogeneous quenches. Performing similar calculations for Renyi QNEC in tractable setups should also lead to non-trivial constraints on the dynamics post-quench.

It seems important to the author to understand the holographic dual of the statement of Renyi QNEC, since it might lead to a generalization of the quantum focussing conjecture. This requires first elucidating the holographic dual of the sandwiched Renyi divergence. The holographic dual of the Renyi entropy was proposed in \cite{Dong:2016fnf} and further studied in \cite{Dong:2016wcf, Nakaguchi:2016zqi, Bianchi:2016xvf, Dong:2017xht, Akers:2018fow}, among others. In particular \cite{Bao:2019aol} studied the holographic dual to the refined Renyi relative entropy. We hope that the techniques of these works might be further developed, and progress made towards a holographic statement and proof for Renyi QNEC.

%%%%%%%%%%%%%%%%%%%%%%%%%%%%%%%%%%%%%%%%%%%%%%%%%%%%%%%%%%%%%%%%%%%%%%%%%%%%%%%%%%%%%
%%%%%%%%%%%%%%%%%%%%%%%%%%%%%%%%%%%%%%%%%%%%%%%%%%%%%%%%%%%%%%%%%%%%%%%%%%%%%%%%%%%%%
\section*{Acknowledgements}

I am grateful to Tanay Kibe for help at a crucial stage in the proof presented in the paper. I would also like to express my gratitude to Suresh Govindarajan, Tanay Kibe, Alok Laddha, Ayan Mukhopadhyay, Rishi Raj and Hareram Swain, for discussions on issues related to this work in and out of the IIT Madras Journal Club. This work was finished at the serene Wits Rural Facility, Hoedspruit, during the 12$^{\rm th}$ Joburg Workshop. I am grateful to Vishnu Jejjala, the organizers, and the University of the Witwatersrand for the Workshop and for the invitation.

%%%%%%%%%%%%%%%%%%%%%%%%%%%%%%%%%%%%%%%%%%%%%%%%%%%%%%%%%%%%%%%%%%%%%%%%%%%%%%%%%%%%%
%%%%%%%%%%%%%%%%%%%%%%%%%%%%%%%%%%%%%%%%%%%%%%%%%%%%%%%%%%%%%%%%%%%%%%%%%%%%%%%%%%%%%
%%%%%%%%%%%%%%%%%%%%%%%%%%%%%%%%%%%%%%%%%%%%%%%%%%%%%%%%%%%%%%%%%%%%%%%%%%%%%%%%%%%%%
\appendix

%%%%%%%%%%%%%%%%%%%%%%%%%%%%%%%%%%%%%%%%%%%%%%%%%%%%%%%%%%%%%%%%%%%%%%%%%%%%%%%%%%%%%
%%%%%%%%%%%%%%%%%%%%%%%%%%%%%%%%%%%%%%%%%%%%%%%%%%%%%%%%%%%%%%%%%%%%%%%%%%%%%%%%%%%%%
%%%%%%%%%%%%%%%%%%%%%%%%%%%%%%%%%%%%%%%%%%%%%%%%%%%%%%%%%%%%%%%%%%%%%%%%%%%%%%%%%%%%%
\section{Free fermion field}		\label{app_fermi_cft}

A Majorana fermion in two-dimensional Minkowski space is described by the action
\begin{equation}
	S = k\int d^2x (-i)\bar\chi\gamma^\mu\del_\mu\chi,
\end{equation}
where $\chi^T = \begin{pmatrix} \chi_1 & \chi_2 \end{pmatrix},  \{\g^\mu,\g^\nu\}=2\h^{\mn}, \bar\chi = \chi^\dagger\g^0$, and $k$ is some normalization factor. We choose
\begin{equation}
	\g^0 = \begin{pmatrix} 0 & \phantom{i}1 \\ 1 & \phantom{i}0 \end{pmatrix}, \qquad \g^1 = \begin{pmatrix} 0 & 1 \\ -1 & 0 \end{pmatrix},
\end{equation}
so that the Majorana condition on $\chi$ implies that both $\chi_1,\chi_2$ are real. After rotating to Euclidean time, $t\to-i\tau$, writing $S_E=-iS$, and defining $z=x-i\tau,\bar z=x+i\tau$, we get
\begin{equation}
	S_E = k\int d\t dx(\psi\bar\del\psi+\bar\psi\del\bar\psi),
\end{equation}
where we have defined $\psi = \sqrt{-i}\chi_1,\bar\psi=\sqrt i\chi_2.$ We focus only on the left-moving chiral field $\psi$ throughout the work, which has the property that
\begin{equation}
	\psi(r)^\dagger = i\psi(r).
\end{equation}
Since $\psi$ is a conformal primary with weight $(h,\bar h)=(\frac12,0)$, we also have
\begin{equation}
	\psi(re^{i\q}) = e^{i\q/2}e^{\q K_P}\psi(r)e^{-\q K_P},
\end{equation}
where $K_P$ is the modular Hamiltonian generating $\q$-rotations for the pencil system.

We choose the normalization factor $k$ to be such that the two-point correlation function is given by
\begin{equation}			\label{eq_fermi_2pt_fn}
	\braket{\psi(z)\psi(w)} = \frac1{z-w} \ .
\end{equation}

%%%%%%%%%%%%%%%%%%%%%%%%%%%%%%%%%%%%%%%%%%%%%%%%%%%%%%%%%%%%%%%%%%%%%%%%%%%%%%%%%%%%%
%%%%%%%%%%%%%%%%%%%%%%%%%%%%%%%%%%%%%%%%%%%%%%%%%%%%%%%%%%%%%%%%%%%%%%%%%%%%%%%%%%%%%
%%%%%%%%%%%%%%%%%%%%%%%%%%%%%%%%%%%%%%%%%%%%%%%%%%%%%%%%%%%%%%%%%%%%%%%%%%%%%%%%%%%%%
\section{Calculating correlation functions}			\label{app_corr_fns}

We want to evaluate
\begin{align}
	\cG(s) =&\ \la (\tilde\r^{(0)})^{-\frac{is}{2\pi}}\cO_{\mu_1\nu_1}(r_1,\q_1)(\tilde\r^{(0)})^{\frac{is}{2\pi}}\ddot\cO_{\mu_2\nu_2}(r_2,\q_2) \ra	\\
		 &\ \qquad + \la (\tilde\r^{(0)})^{-\frac{is}{2\pi}}\dot\cO_{\mu_1\nu_1}(r_1,\q_1)(\tilde\r^{(0)})^{\frac{is}{2\pi}}\dot\cO_{\mu_2\nu_2}(r_2,\q_2) \ra.	\nn
\end{align}
Recalling (\ref{eq_cO_defn}) and that $\tilde\r^{(0)}=\s_P\otimes\tilde\r_A^{(0)}$, this correlation function factorizes into separate correlation functions for the pencil and the auxiliary system,
\begin{equation}
	\cG(s) = \cG_P(s) \cdot \cG_A(s),
\end{equation}
where
\begin{align}
	\cG_P(s)	=&\ \la \s_P^{-\frac{is}{2\pi}}\psi(r_1e^{i\q_1})\s_P^{\frac{is}{2\pi}}\del^2\psi(r_2e^{i\q_2}) \ra_P + \la \s_P^{-\frac{is}{2\pi}}\del\psi(r_1e^{i\q_1})\s_P^{\frac{is}{2\pi}}\del\psi(r_2e^{i\q_2}) \ra_P,
\end{align}
and
\begin{equation}
	\cG_A(s) = \la (\tilde\r_A^{(0)})^{-\frac{is}{2\pi}}E_{\mu_1\nu_1}(\q_1)(\tilde\r_A^{(0)})^{\frac{is}{2\pi}}E_{\mu_2\nu_2}(\q_2)  \ra_A.
\end{equation}
The auxiliary system correlation function can be calculated straightforwardly. For $\q_1>\q_2,$ we get
\begin{align}
	\cG_A(s) =&\ e^{-2\pi K_{\mu_1}}e^{is(K_{\mu_1}-K_{\mu_2})}e^{\q_1(K_{\mu_1}-K_{\nu_1})}e^{\q_2(K_{\mu_2}-K_{\nu_2})}\Tr_a\big[ \ketbra{\mu_1}{\nu_1}\cdot\ketbra{\mu_2}{\nu_2} \big]		\nn		\\
	=&\	e^{-2\pi K_{\mu_1}}e^{(is+\q_{12})K_{\mu_1\mu_2}}\d_{\mu_1\nu_2}\d_{\mu_2\nu_1},
\end{align}
where, as above, $\q_{ij}=\q_i-\q_j$, and also $K_{\mu_i\mu_j}=K_{\mu_i}-K_{\mu_j}$. Similarly, for $\q_2>\q_1$, we have
\begin{equation}
	\cG_A(s) = -e^{-2\pi K_{\mu_2}}e^{(is+\q_{12})K_{\mu_1\mu_2}}\d_{\mu_1\nu_2}\d_{\mu_2\nu_1}.
\end{equation}
where the minus sign appears due to angle-ordering the $E_{\mn}$, which have fermionic statistics. We can combine the above two equations to write for all $\q_1,\q_2$,
\begin{equation}
	\cG_A(s) = {\rm sign}(\q_{12}) e^{-\pi(K_{\mu_1}+K_{\mu_2})}e^{-{\rm sign}(\q_{12})\pi K_{\mu_1\mu_2}}e^{(is+\q_{12})K_{\mu_1\mu_2}}\d_{\mu_1\nu_2}\d_{\mu_2\nu_1}.
\end{equation}

Let us now look at $\cG_P(s)$. Using
\begin{equation}
	\s_P^{-\a}\del^m\psi(re^{i\q})\s_P^\a = e^{i2\pi\a(m+\frac12)}\del^m\psi(re^{i(\q+2\pi\a)}),
\end{equation}
and the fermion two point function (\ref{eq_fermi_2pt_fn}), we get
\begin{equation}
	\cG_P(s) = \frac2{(r_1e^{i\q_1-s}-r_2e^{i\q_2})^3} (e^{-s/2} - e^{-3s/2}).
\end{equation}
Multiplying $\cG_A(s)$ and $\cG_P(s)$, we get
\begin{equation}
	\cG(s) = -{\rm sign}(\q_{12}) 2e^{-\pi(K_{\mu_1}+K_{\mu_2})} e^{-{\rm sign}(\q_{12})\pi K_{\mu_1\mu_2}}e^{\q_{12}K_{\mu_1\mu_2}}\d_{\mu_1\nu_2}\d_{\mu_2\nu_1}\frac{(e^{-s/2} - e^{-3s/2})e^{isK_{\mu_1\mu_2}}}{(r_2e^{i\q_2}-r_1e^{i\q_1-s})^3},
\end{equation}
where we have collected the $s$ dependent terms.

We now want to take a Fourier transform to calculate $G(\w)$,
\begin{equation}
	G(\w) = \frac1{2\pi}\int_{-\I}^\I ds\,e^{-is\w}\cG(s).
\end{equation}
Note that $\cG(s)\to0$ as ${\rm Re}(s)\to\pm\I.$ Also note that
\begin{equation}
	\cG(s+2\pi i) = -e^{-2\pi K_{\mu_1\mu_2}}\cG(s).
\end{equation}
These allow us to write
\begin{equation}
	G(\w) = \frac1{2\pi}\frac1{1+e^{2\pi(\w-K_{\mu_1\mu_2})}}\oint_Cds\,e^{-is\w}\cG(s),
\end{equation}
where $C$ is the closed contour
\begin{equation}
	C:(-\I,\I)\cup(\I,\I+2\pi i)\cup(\I+2\pi i,-\I+2\pi i)\cup(-\I+2\pi i,-\I).
\end{equation}
This contour integral can now be evaluated using the residue theorem, noting the fact that $\cG(s)$ has only one pole inside $C$, given by
\begin{equation}
	s = s_* = \log\frac{r_1}{r_2} + i(\q_{12}+\pi(1-{\rm sign}(\q_{12}))),
\end{equation}
where the extra terms take care of the branches when $\q_{12}<0.$ For $z\in\C$, one can calculate that
\begin{align}
	{\rm Res}\lB \frac{e^{-3s/2}e^{isz}}{(1-re^{i\q-s})^3}, s=s_* \rB =&\ -\frac18{\rm sign}(\q) e^{-\pi z(1-{\rm sign}(\q))}(re^{i\q})^{-\frac32+iz}(1+4z^2),	\\
	{\rm Res}\lB \frac{e^{-s/2}e^{isz}}{(1-re^{i\q-s})^3}, s=s_* \rB =&\ -\frac18{\rm sign}(\q) e^{-\pi z(1-{\rm sign}(\q))}(re^{i\q})^{-\frac12+iz}(-3-8iz+4z^2).
\end{align}
Putting everything together, and using $(1+e^{-2x})^{-1}=e^x\sech x/2,$ we get the following. (Note that we write $z=K_{\mu_1\mu_2}-\w$.)
\begin{align}
	G(\w)	=&\	\frac{i\,e^{\pi(K_{\mu_1\mu_2}-\w)}}{2\cosh\pi(K_{\mu_1\mu_2}-\w)}{\rm Res}\lB e^{-is\w}\cG(s),s=s_* \rB	\nn	\\
	=&\	\frac{i\,e^{\pi z}}{8\cosh\pi z}  e^{-\pi(K_{\mu_1}+K_{\mu_2})} e^{-{\rm sign}(\q_{12}) \pi K_{\mu_1\mu_2}}e^{\q_{12} K_{\mu_1\mu_2}} \d_{\mu_1\nu_2}\d_{\mu_2\nu_1} (r_2e^{i\q_2})^{-3}	\nn	\\
	&\ \quad \times e^{-\pi z(1-{\rm sign}(\q_{12}))} \lb \frac{r_1e^{i\q_1}}{r_2e^{i\q_2}} \rb^{-\frac32+iz} \Big[\lb\frac{r_1e^{i\q_1}}{r_2e^{i\q_2}}\rb(4z^2 - 8iz - 3 ) - (4z^2+1) \Big]	\nn	\\
	=&\ -\frac i8 \d_{\mu_1\nu_2}\d_{\mu_2\nu_1} e^{-\pi (K_{\mu_1}+K_{\mu_2})} e^{-{\rm sign}(\q_{12})\pi\w} (r_1e^{i\q_1})^{-\frac32}(r_2e^{i\q_2})^{-\frac32} \lb\frac{r_1}{r_2}\rb^{iK_{\mu_1\mu_2}}	\nn	\\
	&\ \quad \times \lB Q(z-i)\lb\frac{r_1e^{i\q_1}}{r_2e^{i\q_2}}\rb^{1-i\w} + Q(z)\lb\frac{r_1e^{i\q_1}}{r_2e^{i\q_2}}\rb^{-i\w}\rB,
\end{align}
where we have defined
\begin{equation}
	Q(x) = \frac{4x^2+1}{\cosh\pi x}.
\end{equation}

\providecommand{\href}[2]{#2}\begingroup\raggedright\endgroup

%\bibliography{refs}{}
%\bibliographystyle{JHEP}
%\end{comment}
\end{document}

%% file: custom.tex
\newcommand{\overbar}[1]{\mkern 1.5mu\overline{\mkern-1.5mu#1\mkern-1.5mu}\mkern 1.5mu}

\newcommand{\del}{\partial}

\newcommand{\tr}{\text{tr}\,}
\newcommand{\Tr}{\text{Tr}\,}

\newcommand{\Rmnum}[1]{\expandafter\@slowromancap\romannumeral #1@}

\newcommand{\nn}{\nonumber}

\newcommand{\be}{\begin{equation}}
\newcommand{\ee}{\end{equation}}

\newcommand{\lB}{\left [}
\newcommand{\rB}{\right ]}
\newcommand{\lb}{\left (}
\newcommand{\rb}{\right )}
\newcommand{\lbr}{\left \{}						%% Left braces
\newcommand{\rbr}{\right \}}					%% Right braces
\newcommand{\ld}{\left.}				        %% Left  continuation of bracketed expression.
\newcommand{\rv}{\right\vert}
\newcommand{\la}{\left\langle}
\newcommand{\ra}{\right\rangle}
\renewcommand{\a}{\alpha}	%%% Redefinition
\renewcommand{\d}{\delta}	%%% Redefinition

\newcommand{\f}{\phi}

\newcommand{\g}{\gamma}
\newcommand{\h}{\eta}

\renewcommand{\k}{\kappa}	%%% Redefinition
\renewcommand{\l}{\lambda}	%%% Redefinition
\newcommand{\mn}{\mu\nu}
\newcommand{\q}{\theta}	
	
\renewcommand{\r}{\rho}		%%% Redefinition
\newcommand{\s}{\sigma}

\renewcommand{\t}{\tau}		%%% Redefinition
\newcommand{\w}{\omega}

\newcommand{\A}{\forall}

\newcommand{\D}{\Delta}
\newcommand{\F}{\Phi}

\newcommand{\I}{\infty}
\newcommand{\W}{\Omega}

\renewcommand{\S}{\Sigma}	%%% Redefinition

\newcommand{\C}{\mathbb{C}}
\newcommand{\R}{\mathbb{R}}
\newcommand{\Z}{\mathbb{Z}}

\newcommand{\cA}{\mathcal{A}}

\newcommand{\cF}{\mathcal{F}}
\newcommand{\cG}{\mathcal{G}}
\newcommand{\cH}{\mathcal{H}}

\newcommand{\cO}{\mathcal{O}}

\newcommand{\sA}{\mathscr{A}}